# Cascaded Parametric Amplification for Highly Efficient Terahertz Generation

Koustuban Ravi,[1,3,4*] Michael Hemmer,[1] Giovanni Cirmi,[1,2] Fabian Reichert,[2] Damian N. Schimpf,[1,2] Oliver D. Mücke,[1] and Franz X. Kärtner[1,2,4]

[1]*Center for Free-Electron Laser Science, Deutsches Elektronen-Synchrotron (DESY), Hamburg, Germany*
[2]*Department of Physics and the Hamburg Center for Ultrafast Imaging, University of Hamburg, Germany*
[3]*Department of Electrical Engineering and Computer Science, Massachusetts Institute of Technology, Cambridge, MA, USA*
[4]*Research Laboratory of Electronics, Massachusetts Institute of Technology, Cambridge, MA, USA*
*Corresponding author: koust@mit.edu*

**A highly efficient, practical approach to high-energy terahertz (THz) generation based on spectrally cascaded optical parametric amplification (THz-COPA) is introduced. The THz wave initially generated by difference frequency generation between a strong narrowband optical pump and optical seed (0.1-10% of pump energy) kick-starts a repeated or *cascaded* energy down-conversion of pump photons. This helps to greatly surpass the quantum-defect efficiency and results in exponential growth of THz energy over crystal length. In cryogenically cooled periodically poled lithium niobate, energy conversion efficiencies >8% for 100 ps pulses are predicted. The calculations account for cascading effects, absorption, dispersion and laser-induced damage. Due to the coupled nonlinear interaction of multiple triplets of waves, THz-COPA exhibits physics distinct from conventional three-wave mixing parametric amplifiers. This in turn governs optimal phase-matching conditions, evolution of optical spectra as well as limitations of the nonlinear process.**

DOI

Multi-cycle or narrowband terahertz (THz) frequency sources with simultaneously high pulse energies (>10 mJ) and peak fields (>100 MV/m) in the frequency range between 0.1 and 1 THz are of great interest for compact particle acceleration [1], coherent X-ray generation [3], linear and nonlinear spectroscopy and radar applications. Such a class of THz sources has the potential to achieve considerable reduction in the cost and size of current accelerators to enable unprecedented modalities in biomedical imaging, therapy and protein structure determination [2]. Among existing THz generation methods, photoconductive switches can be efficient [3], but are challenging to scale to high pulse energies.

Vacuum electronic devices such as gyrotrons [4] are limited in their frequency of operation or peak powers while free-electron lasers are expensive large-scale facilities [5].

Due to the rapid increase in laser pulse energies produced by solid-state laser sources, laser-driven narrowband THz generation methods based on difference-frequency generation (DFG) are attractive. However, an issue that needs to be addressed is the realization of high optical-to-THz energy conversion efficiencies (or conversion efficiencies), particularly at high pump energies. To generate tens of millijoules (mJs) of THz energy from Joule-class lasers, conversion efficiencies >>1% are necessary.

Previous work on multi-cycle THz generation demonstrated conversion efficiencies in gallium arsenide (GaAs) of $10^{-4}$ [6, 7], in gallium phosphide $10^{-6}$ [8], and organic materials $10^{-5}$ [9]. In lithium niobate (LN), multi-cycle THz generation by interfering chirped and delayed copies of a pulse with tilted-pulse-fronts (TPF) was demonstrated [10]. However, TPFs have limitations due to group-velocity dispersion (GVD) induced by angular dispersion [11]. Such issues were circumvented by optical rectification in cryogenically cooled periodically poled lithium niobate (PPLN) crystals, but the conversion efficiency was only 0.1 % at 0.5 THz [12] due to large walk-off between the optical pump pulse and generated THz radiation. In order to obtain significantly larger conversion efficiencies, a feasible method which preserves phase-matching and minimizes loss to enable repeated down conversion of optical pump photons to THz photons (or cascading effects) is required [13].

Here, we introduce a practical approach based on narrowband (hundreds of ps transform-limited (TL) duration) pulses utilizing cascading effects. It involves using a strong narrowband optical pump pulse along with a narrowband seed, shifted in frequency from the pump by the desired THz frequency and with significantly lower pulse energy (~0.1-10% of pump). The initial DFG between these pulses generates narrowband THz radiation, which then drives dramatic *cascading* of the optical spectrum resulting in a shift, broadening of the optical spectrum and *exponential* THz energy growth over the crystal length. Since the rate of exponential growth is *parameter* dependent i.e. on pump intensity, THz frequency etc., the approach is termed THz cascaded optical parametric amplification (THz-COPA) [14]. Conversion efficiencies >8% in cryogenically cooled PPLN crystals are predicted. For pump pulse energy of 1J, the corresponding seed energy would span the mJ-100mJ range. Such sources are readily available at kHz repetition rates with 1-μm laser technology [15]. Furthermore, the required crystal apertures for Joule-level pulses with durations of hundreds of ps are < 1 cm$^2$, which is attainable with existing technology [16].

Previously, THz seeded parametric amplification [17] was demonstrated in a non-collinear geometry in LN at room temperature. Moreover, optically seeded approaches in parametric oscillator configurations [18, 19] were demonstrated in non-collinear geometries at room temperature. Cavity-based configurations limit the number of cascades that the pump may undergo, while THz seeded approaches suffer from the limited power levels at THz frequencies available from compact seed sources.

The THz-COPA case analyzed employs cryogenic cooling and collinear PPLN geometries in a single-pass configuration to realize long interaction lengths and enable significant cascading. The use of long pump pulses is advantageous as it alleviates laser-induced damage, walk-off between optical and THz radiation and nonlinear phase accumulation. In combination, this results in the realization of high conversion efficiencies.

It is worthwhile noting that the *serial combination* of optical parametric amplifiers (OPAs), has been referred to as COPA in the past [20]. However, that work is still a conventional three-wave mixing process. The process described in this paper involves the simultaneous nonlinear interaction of multiple triplets of waves. We thus discuss the distinctive evolution of optical spectra in THz-COPA and deduce preferred phase-matching conditions and resulting limitations. An example of a distinctive feature is that the pump is not required to be the highest frequency involved. This is because, the initially produced THz by DFG drives the generation of adjacent red-shifted frequency lines leading to further THz energy growth.

For the case under consideration, the optical spectrum consists of narrowband frequency distributions centered on various frequencies $f_m = f_0 + m f_{THz}$, where $f_0$ is the optical pump frequency, $m$ is an integer and $f_{THz}$ is the generated THz frequency. Since the frequency distributions are narrowband, we may approximate them as discrete lines (or modes) in frequency space. The discrete modes are characterized by amplitudes $E_m(z) = A_m(z)e^{-jk_m z}$ where, $A_m$ is the envelope, $k_m$ is the wave number of mode $m$. The evolution of $A_m(z)$ along the crystal length $z$ is given by

$$\frac{dA_m}{dz} = \frac{-j2f_m\chi_0^{(2)}}{cn(f_m)}[A_{m+1}A^*{}_{THz}\, e^{-j(k_{m+1}-k_{THz}-k_m+2\pi\Lambda^{-1})z} \\ + A_{m-1}A_{THz}e^{-j(k_{m-1}+k_{THz}-k_m+2\pi\Lambda^{-1})z}] \quad (1)$$

The first term on the right-hand side (RHS) of Eq. (1) corresponds to DFG between the (m+1)$^{th}$ optical mode and the THz mode $A_{THz}(z)$. The second term on the RHS of Eq. (1) corresponds to sum-frequency generation (SFG) between the THz and (m-1)$^{th}$ optical mode. $\Lambda$ is the PPLN period, $\chi_0^{(2)}$ is the second-order susceptibility and $n(f_m)$ is refractive index at frequency $f_m$. Similarly, the evolution of the THz mode $A_{THz}(z)$ is given by

$$\frac{dA_{THz}}{dz} = \frac{-\alpha}{2}A_{THz} - \frac{j2f_{THz}\chi_0^{(2)}}{cn(f_{THz})}\sum_{m=1}^{N-1}A_{m+1}A_m^* e^{-j(k_{m+1}-k_m-k_{THz}+2\pi\Lambda^{-1})z} \quad (2)$$

The first term on the RHS of Eq. (2) corresponds to THz absorption, while the second term corresponds to THz generation by the sum of all possible DFG processes between optical modes. We thus see how Eqs.(1)-(2) involve the coupled nonlinear interaction of multiple-triplets ($A_{m+1}$, $A_m$, $A_{THz}$) of waves. Only the first quasi-phase-matched order THz wave is considered since the generation efficiency decreases as $p^{-2}$ while absorption increases for higher $p^{th}$ orders. Equations (1)-(2) include dispersion in the optical region and THz absorption. For the relatively low intensities associated with the long pulse durations specific to this problem, self-phase modulation (SPM) effects may be omitted. In general, the conversion efficiency in the discrete case is the ratio of the total spectral intensities

$$\eta(z) = \frac{n(f_{THz})|A_{THz}(f_{THz},z)|^2}{\sum_m n(f_m)|A_m(f_m,0)|^2} \quad (3)$$

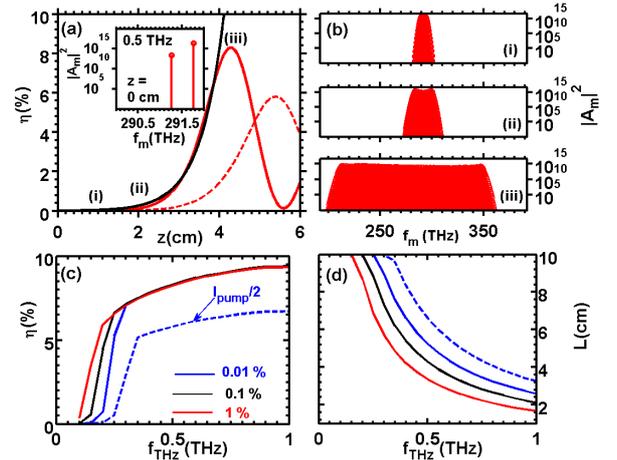

Fig.1 (a) Inset: Pump at 291.76 THz with red-shifted seed with 0.1% of pump energy. Conversion efficiency ($\eta$) versus crystal length (red-solid) is fit to hyperbolic sine (~exponential) growth (black). At half the pump intensity, η grows slower with comparable maximum. Decline in η is due to alteration of phase-matching caused by spectral shift; (b) optical spectra at locations (i)-(iii) indicated in panel (a); (c) maximum η as a function of seed level (0.01 to 1% of pump energy) and THz frequency at different pump intensities; (d) optimal crystal lengths corresponding to parameters panel (c).

The nonlinear material under consideration is magnesium oxide doped (5% mol.) congruent LN. The refractive indices and absorption coefficients are obtained from [21, 22, 23]. The optical modes span a frequency bandwidth from $f_m$=100 to $f_m$=500 THz. Equations (1)-(3) are solved numerically via a 4$^{th}$ order Runge-Kutta method with spatial resolution $\Delta z$= 0.5 μm. The second-order susceptibility $\chi_0^{(2)}$ is 336 pm/V. The maximum fluence, limited by laser-induced damage, is phenomenologically related to the 1/$e^2$ pulse duration $\tau$ as $F_d = 8.5(\tau/10ns)^{1/2}$ Jcm$^{-2}$. A PPLN crystal phase-matched for generation of 0.5 THz radiation with poling period $\Lambda = 238$ μm at a cryogenic temperature of 77K is assumed. As shown in the inset of Fig. 1a, the pump mode corresponding to a pulse with TL duration $\tau=100$ ps was located at $f_0=291.76$ THz (1028.2 nm) with pump intensity $F_d\left(\tau\sqrt{\pi/2}\right)^{-1}$. The seed pulse, also corresponding to a TL duration $\tau=100$ ps, is red-shifted by 0.5 THz with respect to the pump with 0.1 % of the pump intensity. Assuming perfect spatial overlap of seed and pump, the seed will contain 0.1% of the pump energy.

In Fig. 1a, the conversion efficiency $\eta$ as a function of length for the aforementioned parameters is plotted (Fig. 1a, red-solid). For $z$ up to 4 cm, the growth of $\eta$ follows a sine hyperbolic (~exponential) fit (Fig. 1a, black). For $z > 4$ cm, $\eta$ drops drastically before ascending again at ~5.6 cm. A second curve (Fig.1a, red-dashed) is plotted for half the maximum pump intensity. The behavior and peak efficiency is similar apart from a delayed onset of growth (or lower gain), similar to conventional OPAs. In Fig. 1b, the optical spectra at values of $z$ indicated by labels (i)-(iii) for the case of maximum pump intensity (Fig. 1a, red-solid) are shown. At (i) $z = 1$ cm and (ii) $z = 2$ cm, before the onset of exponential growth, a modulation of the optical spectrum with roughly symmetric red (DFG) and blue-shift (SFG) is observed. During this initial stage, THz growth is not significant due to comparable rates of SFG and DFG. However, as THz radiation builds up significantly, a red-shift is preferred as evident at (iii) $z = 4$ cm and is accompanied by high conversion efficiencies in excess of 8 %. These optimal crystal lengths (4-5 cm) are within the limits of existing technology. Thus, using THz-COPA, THz pulses with tens of mJs of pulse energy may be obtained with existing laser and crystal fabrication technology. The decline and subsequent oscillation of $\eta$ in Fig. 1a is attributed to the change in phase-matching conditions caused by the frequency shift and spectral broadening of the optical spectrum. This may be remedied by re-using the optical spectrum for THz generation in a subsequent stage or varying the PPLN period along the crystal length.

In Fig. 1c, the maximum conversion efficiency achieved over PPLN crystal lengths <10 cm for various THz frequencies and seed levels are depicted. The rate of energy growth for lower THz frequencies is lower, in line with expectations from conventional OPAs with a gain proportional to $(f_0 f_{THz})^{1/2}$. Therefore, for a given seed level, frequencies beyond a threshold value are able to reach maximum conversion efficiency within the 10 cm length. Naturally, this threshold value decreases with increasing seed level or peak pump intensity. The corresponding optimal crystal lengths are plotted in Fig. 1d. Consistent with expectations, the optimal crystal lengths are shorter for larger THz frequencies, and higher seed levels and pump intensities (Fig. 1d, blue-dashed). For longer pulse durations $\tau$, the peak pump intensity reduces due to damage restrictions. In such cases, longer crystals or larger seed levels may be necessary.

In Fig. 2, we examine the evolution of optical spectra for various cases of phase-mismatch $\Delta k_m$ (Fig. 2a) to shed light on the physics of THz-COPA. Absorptive effects are switched off to not obfuscate the essential physics. In Fig. 2b, the conversion efficiency $\eta$ as a function of $z$ for cases (i)-(iii) of $\Delta k_m$ are shown. Case (i) maps the realistic situation for the case of PPLN simulated in Fig. 1, i.e., $\Delta k_m = k_{m+1} - k_m - k_{THz} + 2\pi\Lambda^{-1}$. For this case, we see that $\Delta k_m$ is almost symmetric for frequencies red and blue-shifted with respect to the pump frequency $f_0$ (Fig. 2a, blue). The corresponding $\eta$ versus $z$ curve (Fig. 2b, blue) resembles the red-solid curve in Fig. 1a.

However, the values of $\eta$ are larger and an oscillation rather than decline is observed for $z > 4$ cm since absorption is neglected. In Fig. 2c, the optical spectrum as a function of $z$ is plotted for case (i). Here, an initial modulation of the optical spectrum, resulting in a large number of side-bands, blue and red-shifted with respect to $f_0$ is observed. Each SFG process consumes a THz photon, whereas each DFG process generates a THz photon. Consequently, during this initial stage, there is little growth in the efficiency of THz radiation. In contrast to conventional three-wave mixing OPAs, this behavior corresponds to a large ensemble of coupled three-wave ($A_{m+1}$, $A_m$, $A_{THz}$) mixing processes. This is made possible by the fact, that THz frequencies are small in comparison to the phase-matching bandwidth of the material. Due to this competition between SFG and DFG, THz seeding in PPLN's may lead to initial THz attenuation before producing similar exponential growth.

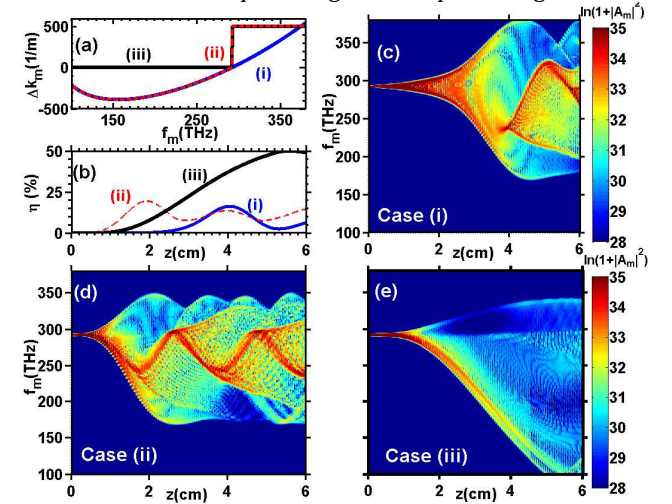

Fig.2 Role of phase-mismatch in THz-COPA: (a) Cases (i)-(iii) of phase-mismatch $\Delta k_m$; (b) corresponding conversion efficiencies for $\Delta k_m$ cases (i)-(iii); (c) evolution of optical spectrum for case (i) showing initial modulation behavior; (d) evolution of optical spectrum for case (ii) showing predominant red-shift due to large phase-mismatch for blue-shifted frequencies ($\Delta k_m=500 m^{-1}$ for $f_m>f_0$); (e) evolution of the optical spectrum for case (iii) shows steady red-shift due to preservation of phase-matching conditions upon red-shift ($\Delta k_m=0$ for $f_m<f_0$).

After initial symmetric broadening, the process shows a preferential red-shift due to marginally larger dispersion for frequencies larger than $f_0$ as seen in case (i) (Fig. 2a, blue). Correspondingly, the THz begins to grow exponentially (Fig. 2b, blue). We confirm our hypothesis by defining $\Delta k_m = 500 m^{-1}$ for $f_m>f_0$, while leaving it unchanged for $f_m<f_0$ in case (ii) (Fig. 2a, red-dashed). Correspondingly, the blue-shift in the spectrum is negligible (Fig. 2d), resulting in an earlier onset of exponential THz growth (Fig. 2b, red). However, for cases (i)-(ii), there is a saturation and oscillation of conversion efficiency beyond a certain length. This is attributed to the change in phase-matching conditions due to the shift in the optical spectrum by cascading. To test this, we define $\Delta k_m = 0$ for $f_m<f_0$ and retain $\Delta k_m=500 m^{-1}$ for $f_m>f_0$ in case (iii) (Fig. 2a, black). We then see that the conversion efficiency does not saturate but continues to grow over a longer portion of the crystal length (Fig. 2b, black). The optical spectrum for case (iii) exhibits a steady and predominant red-shift as delineated in Fig. 2e. The peak pump intensity for case (iii) is reduced to half the value of cases (i)-(ii), so that maximum conversion efficiency is attained at $z=6$ cm.

The simulations in Fig. 2 thus illustrate the need for discriminating between blue and red-shift. As pointed out earlier, this behavior is markedly different from conventional OPAs, where the large frequency differences in the optical spectral region between pump and signal, typically restrict the behavior to a three-wave mixing process. For instance, in the multi-triplet case, in the absence of any dispersion in the material, i.e., $\Delta k_m=0$, a mere oscillation of conversion efficiency as shown in Fig. 3a is observed. In Fig. 3b, the corresponding evolution of the optical spectrum is displayed. Symmetric broadening or modulation of the optical spectrum; equally red and blue-shifted with respect to the pump frequency $f_0$ is observed.

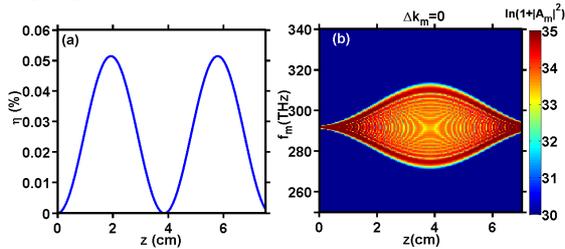

Fig. 3(a) For perfect phase-matching, i.e., $\Delta k_m=0$, conversion efficiency shows oscillatory behavior with respect to length. (b) Optical spectrum shows symmetric broadening due to modulation.

A simple analysis sheds light on this behavior: Suppose, only the pump ($A_0$), two side-bands $f_0 \pm f_{THz}$ ($A_{\pm 1}$) and the THz wave ($A_{THz}$) from Eqs. (1)-(2) are considered in the undepleted limit of the pump, we obtain oscillatory behavior for the side-bands and THz wave in the case of perfect phase matching.

$$A_{THz}(z) = A_{THz,\max}\sin(\kappa z),\ \kappa = 2\sqrt{2}\chi_0^{(2)}|A_0|f_{THz}c^{-1}(n(f_0)n(f_{THz}))^{-1/2} \quad (4)$$
$$A_{\pm 1}(z) = A_{\pm}(0) + j4(c\kappa n(f_0))^{-1}(f_0 \pm f_{THz})\chi_0^{(2)}A_0 A_{THz,\max}\sin^2(\kappa z/2)$$

On the contrary, exponential growth is obtained in the three-wave case. Another example of interesting behavior is indifference to the location of the seed with respect to the pump. In conventional OPAs for the difference frequency, the pump has the highest frequency compared to signal and idler. However, due to the multi-triplet nature of the THz-COPA, the initial spectrum rapidly washes out leading to exponential THz growth as shown in Fig. 4.

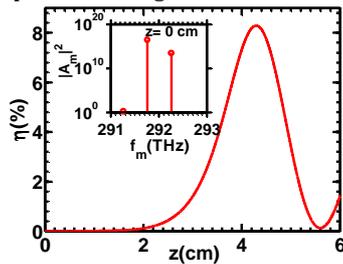

Fig.4. Indifference to seed location: Contrary to conventional OPAs, the THz-COPA shows exponential growth even if the seed is at higher frequency relative to the pump (inset). The seed contains 0.1% of the pump energy.

A practical approach to high-energy THz generation based on cascaded optical parametric amplification (THz-COPA) was introduced. Starting with a strong narrowband optical pump and significantly weaker optical seed pulse, extensive frequency shifting due to cascading occurs which results in high conversion efficiencies at the 10 % level in cryogenically cooled PPLNs. Further improvements may be obtained by multi-staging or varying the PPLN period along the crystal. The THz-COPA process exhibits distinctively different physics from conventional OPAs. This influences the optimal pulse formats and phase-matching conditions. The approach paves the way for extremely high energy THz sources using laser technology.

**Funding** European Research Council (grant n. 609920); "The Hamburg Centre for Ultrafast Imaging- Structure, Dynamics and Control of Matter at the Atomic Scale" of the Deutsche Forschungsgemeinschaft (grant EXC 1074).

**References**

[1] L. J. Wong, A. Fallahi and F. X. Kärtner, "Compact electron acceleration and bunch compression in THz waveguides," *Optics Express,* vol. 21, no. 8, pp. 9792-9806, 2013.

[2] F. X. Kärtner, "AXSIS:Exploring the frontiers in attosecond X-ray Science," *Nuclear Instruments and Methods in Physics Research A,* http://dx.doi.org/10.1016/j.nima.2016.02.080 (2016).

[3] N. Yardimici, S.-H. Yang, C. W. Berry and M. Jarrahi, "High-Power Terahertz Generation Using Large-Area Plasmonic Photoconductive emitters," *IEEE Transactions on Terahertz Science and Technology,* vol. 5, no. 2, pp. 223-229, 2015.

[4] S. Gold and G. Nusinovich, "Review of high-power microwave source research," *Review of Scientific Instruments,* vol. 68, no. 11, p. 3945, 1997.

[5] G. P. Gallerano and S. Biedron, "Overview of terahertz radiation sources," *Proceedings of the FEL conference,* p. 216, 2004.

[6] K. L. Vodopyanov, "Optical THz-wave generation with periodically-inverted GaAs," *Laser and photonics reviews,* vol. 2, pp. 11-25, 2008.

[7] K. Vodopyanov, M. Fejer, X. Yu, J. S. Harris, Y. S. Lee , W. C. Hurlbut, V. G. Kozlov, D. Bliss and C. Lynch, "Terahertz-wave generation in quasi-phase-matched GaAs," *Applied Physics Letters,* vol. 89, p. 141119, 2006.

[8] G. Kitaeva, "Terahertz generation by means of optical lasers," *Laser Physics Letters,* vol. 5, p. 559, 2008.

[9] J. Lu, H. Hwang, X. Li, S.-H. Lee, O. Pil-Kwon and K. A. Nelson, "Tunable multi-cycle THz generation in organic crystal HMQ-TMS," *Optics Express,* vol. 23, pp. 22723-22729, 2015.

[10] Z. Chen, X. Zhou, C. Werley and K. A. Nelson, "Generation of high power tunable multicycle terahertz pulses," *Applied Physics Letters,* vol. 99, p. 071102, 2011.

[11] K. Ravi, W. Huang, S. Carbajo, X. Wu and F. X. Kärtner, "Limitations to THz generation by optical rectification using tilted pulse fronts," *Optics Express,* vol. 22, no. 17, pp. 20239-20251, 2014.

[12] S. Carbajo, J. Schulte, W. Xiaojun, K. Ravi, D. Schimpf and F. Kärtner, "Efficient narrowband terahertz generation in cryogenically cooled periodically poled lithium niobate," *Optics Letters,* vol. 40, pp. 5762-5765, 2015.

[13] M. C. Golomb, "Cascaded nonlinear difference frequency generation of enhanced wave production," *Optics Letters,* vol. 29, p. 2046, 2004.

[14] F. X. Kärtner, K. Ravi, D. Schimpf, G. Cirmi, M. Hemmer, O. Muecke, A. Fallahi, N. Matlis, F. Reichert, G. M. Rossi and L. Zapata, "Method and apparatus for generating THz radiation". European Patent 16000684.7, 2016.

[15] L. Zapata, H. Lin, A.-L. Calendron, H. Cankaya, M. Hemmer, F. Reichert, W. R. Huang, E. Granados, K.-H. Hong and F. X. Kaertner, "Cryogenic Yb:YAG composite thin-disk for high energy and average power amplifiers," *Optics Letters,* vol. 40, pp. 2610-2613, 2015.

[16] T. Taira and H. Ishizuki, "Improvement of laser-beam distortion in large-aperture PPMgLN device by using X-axis Czochralski-grown crystal," *Optics Express,* vol. 22, p. 19668, 2014.

[17] S. R. Tripathi, Y. Taira, S. Hayashi, K. Nawata, K. Murate, H. Minamide and K. Kawase, "Terahertz wave parmetric amplifier," *Optics Letters,* vol. 39, no. 6, p. 1649, 2014.

[18] D. Molter, M. Theuer and R. Beigang, "Nanosecond terahertz optical parametric oscillator with a novel quasi phase matching scheme in lithium niobate," *Optics Express,* vol. 17, p. 6623, 2009.

[19] D. Walsh, P. Browne, M. Hunn and C. Rae, "Intracavity parametric generation of nanosecond terahertz radiation using quasi-phase-matching," *Optics Express,* vol. 18, p. 13951, 2010.

[20] I. Jovanovic, C. P. Barty, C. Haefner and B. Wattellier, "Optical switching and contrast enhancement in intense laser systems by cascaded optical parametric amplification," *Optics Letters,* vol. 31, p. 787, 2006.

[21] D. H. Jundt, "Temperature-dependent Sellmeier equation for the index of refraction, ne, in congruent lithium niobate," *Optics Letters,* vol. 22, pp. 1553-1555, 1997.

[22] L. Palfalvi, J. Hebling, J. Kuhl, A. Peter and K. Polgar, "Temperature dependence of the absorption and refraction of Mg-doped congruent and stoichiometric lithium niobate in the THz range," *Journal of Applied Physics,* vol. 97, p. 123505, 2005.

[23] J. A. Fulop, L. Palfalvi, M. Hoffmann and J. Hebling, "Towards generation of mJ-level ultrashort THz pulses by optical rectification," *Optics Express,* vol. 19, pp. 15090-15097, 2011.